\documentclass[%
 reprint,
superscriptaddress,
 amsmath,amssymb,
 aps, prl, nofootinbib,
]{revtex4-2}
\usepackage{tabularx}
\usepackage{orcidlink}
\usepackage{multirow}
\usepackage{graphicx}
\usepackage{dcolumn}
\usepackage{bm}
\usepackage{bbding}
\usepackage[mathlines]{lineno}
\usepackage{appendix}
\usepackage{url}
\usepackage{algorithm}
\usepackage{algpseudocode}

\renewenvironment{algorithm*}[1][t]
  {\begin{figure*}[#1]
   \begin{minipage}{\linewidth}
   \begin{algorithm}[H]}
  {\end{algorithm}
   \end{minipage}
   \end{figure*}}

\begin{document}

\preprint{APS/123-QED}

\title{Unsupervised Neural Operators for Rotating Relativistic Stars: High-Fidelity Equation of State Inference and Pulsar Spin Evolution at Scale}
\author{Anarya Ray\,\orcidlink{0000-0002-7322-4748
}}
\email{anarya.ray@northwestern.edu} \affiliation{Center for Interdisciplinary Exploration and Research in Astrophysics (CIERA), Northwestern University, 1800 Sherman Ave,
Evanston, IL 60201, USA}
\affiliation{NSF-Simons AI Institute for the Sky (SkAI), 172 E. Chestnut Street, Chicago, IL 60611, USA}

\date{\today}

\begin{abstract}
Observed populations of rotating neutron stars~(NSs) with millisecond-order periods are powerful probes of fundamental physics. Mass-radius measurements inform poorly constrained properties of degenerate nuclear matter, whereas high-fidelity population synthesis connects unknowns in NS formation and binary evolution to large radio-timing surveys. However, repeated structure calculations are prohibitive due to the computational cost of solving the highly coupled partial differential equations of axisymmetric general relativistic hydrostatic equilibrium that govern rotating NS interiors. Non-rotating calculations or approximate rotational corrections can bias equation of state inference and limit the fidelity of population synthesis in tracing the evolutionary history of progenitors and propagating equation of state uncertainties. In this \textit{Letter}, we develop physics-informed neural operators that can solve the axisymmetric problem with high accuracy and negligible computational cost. \textit{Our solvers are trained in a fully unsupervised setting, with the training objective informed solely by the partial differential equations, and achieve a $10^3\times$ speed-up with negligible accuracy loss}. These solvers enable scalable EoS inference that self-consistently accounts for rotation and repeated stellar-structure calculations during binary evolution, making spin predictions that account for rotational structure and EoS uncertainties computationally accessible to population synthesis.
\end{abstract}
\maketitle



\emph{Introduction} --- Neutron stars~(NSs) are natural laboratories for dense matter physics~\cite{LattimerPrakash2007}. Observable structural properties are governed by general relativistic hydrostatic equilibrium, with the NSs' self-gravity supported by neutron degeneracy pressure~\cite{LattimerPrakash2007}. The equation of state~(EoS: $p=p(\epsilon)$), a functional relationship between energy density~$(\epsilon)$ and degeneracy pressure~$(p)$, determines unique solutions which predict universal mass-radius relationships, with theoretical models subject to uncertainties in the treatments of low-energy strong nuclear interactions~\cite{LattimerPrakash2007,Hebeler2013}. Mass-radius measurements from astronomical observations therefore facilitate crucial empirical probes of low-energy nuclear theory beyond the reach of terrestrial experiments~\cite{Hebeler2013,OzelFreire2016,Ozel2016,Essick2020Chiral,Dietrich2020}. The universality of the underlying EoS enables increasingly precise constraints from a growing number of independent observations across multiple observational probes~\cite{Fasano2019,Miller2020EOS,Landry2020,Raaijmakers2020,Capano2020,Dietrich2020,Raaijmakers2021,Legred2021,Biswas2022,Huth2022,Pang2023}. The same stellar structure also governs the response of NSs to accretion, connecting dense matter physics to the spin evolution~\cite{Cook1994Recycling,Burderi1999,Lavagetto2004,Bejger2011,Fortin2016,Holgado2021} and observable population properties~\cite{Possenti1998,Possenti1999,GittinsAndersson2019,Chattopadhyay2021,Chattopadhyay2026,Sautron2026} of rotating NSs in binaries.

Traditional electromagnetic~(EM) probes---X-ray timing, burst spectroscopy, and thermal emission from quiescent low-mass X-ray binaries, often combined with radio timing---yield independent mass-radius measurements~\cite{OzelFreire2016,Riley2019}, often susceptible to systematic biases~\cite{MillerLamb2016}. More recent gravitational-wave~(GW) and multimessenger observations of NS-containing compact binary mergers provide a cleaner complementary probe~\cite{Abbott2018EOS,Ghosh2021,Ray2023,GolombTalbot2022}, enabling joint empirical EoS constraints~\cite{Dietrich2020,Landry2020,Wagner2025RIFT,Vilkha2024RIFT,OShaughnessyZhu2026}. Complementarily, the masses, spins, and orbital properties of radio pulsar populations probe NS formation and binary evolution through comparisons with population synthesis~\cite{Kiel2008,Oslowski2011,VignaGomez2018,Chattopadhyay2020,Sgalletta2023,Chattopadhyay2026,Sautron2026}.

Both investigations necessitate repeated structure calculations, solving the differential equations of hydrostatic equilibrium for a given pressure-density table to obtain a mass-radius relationship. Constraining theoretical models from mass-radius observations relies on Bayesian hierarchical inference~\cite{Riley2018EOS,Miller2020EOS,Landry2020, Ray2023}, with Monte Carlo sampling requiring multiple solver calls to obtain empirical credible intervals in the pressure-density plane from independent EoS-agnostic measurements across multi-probe datasets~\cite{Riley2018EOS,Raaijmakers2021,Legred2021}. Similarly, EoS-dependent stellar structure affects accretion-driven spin-up~\cite{Cook1994Recycling,Burderi1999,Holgado2021}, a key process connecting binary evolution to observable millisecond pulsar~(MSP) populations~\cite{Possenti1998,Possenti1999,Chattopadhyay2026}. Tracing this spin evolution throughout mass transfer requires repeated structure calculations along each evolutionary trajectory~\cite{Lavagetto2004,Bejger2011,Fortin2016}, multiplying the cost of synthesizing populations and propagating EoS uncertainties. Scalable solvers are hence crucial for both EoS inference and population synthesis of accreting NSs. 

The scalability of numerical solvers depends on the complexity of the differential equations. In the spherical~(Tolman--Oppenheimer--Volkoff, TOV) limit, these reduce to an ODE with fast solvers and inverse methods available~\cite{Tolman1939,OppenheimerVolkoff1939,Lindblom1992}. However, observed millisecond rotation periods motivates axisymmetric equilibrium descriptions~\cite{OzelFreire2016,Riley2019}, involving five coupled equations, including three elliptic PDEs, whose costlier solvers can render Monte Carlo EoS exploration and repeated calculations across synthetic populations computationally prohibitive~\cite{StergioulasFriedman1995,Gourgoulhon2010}. Resorting to the TOV limit or finite-order rotational corrections~\citep{Hartle1967,HartleThorne1968} may introduce systematic biases exceeding future measurement uncertainties~\cite{Baubock2015}.

In this \textit{Letter}, we propose accurate and scalable solvers based on physics-informed neural operators~\cite{Wang2021PIDeepONet,Li2021PINO,Bezerra2026}, present EoS analyses that self-consistently account for rotation, and investigate the effects of rotational structure on pulsar spin evolution in binary systems. Complementing tabulated equilibria and supervised surrogates~\cite{Lavagetto2004,Fortin2016,Liu2026,Papigkiotis2025}, \textbf{our solvers learn continuous (mesh-free) stellar profiles from physics alone}, retaining explicit dependence on the $p(\epsilon)$ representation of the EoS with a one-time training across the EoS prior. We show that our solvers are \textbf{three orders of magnitude faster than numerical implementations with negligible loss in accuracy} and mitigate substantial EoS inference biases that arise from neglecting rotation or using second-order corrections~\citep{Hartle1967,HartleThorne1968}. The same framework enables repeated stellar-structure calculations during binary evolution, \textbf{making spin predictions that account for rotational structure and EoS uncertainties computationally accessible to population synthesis}. Below, we present our novel methodology and analysis, preceded by a brief introduction to rotating relativistic stars, and conclude thereafter with a discussion on relevance, impact, and ongoing investigations. 

\emph{Uniformly Rotating Relativistic Stars} --- Rotating NSs are governed by the axisymmetric equations of general Relativistic Hydrostatic Equilibrium, which can be solved given the functional relationship between neutron degeneracy pressure and energy density to uniquely determine the stellar profile for any chosen central density and angular speed~\cite{Gourgoulhon2010}. Under axisymmetry and uniform rotation,  the relevant components of the metric and energy-momentum tensors can be expressed as~\cite{Komatsu1989,Gourgoulhon2010}:
\begin{equation}
\begin{aligned}
ds^2 &= -e^{\gamma+\rho}dt^2
       +e^{2\alpha}(dr^2+r^2d\theta^2)\\
     &\quad +e^{\gamma-\rho}r^2\sin^2\theta
       (d\phi-\omega\,dt)^2,\\
T^{\mu\nu} &= (\epsilon+p)u^\mu u^\nu+p\,g^{\mu\nu},\\
u^\mu &= \frac{e^{-(\gamma+\rho)/2}}{\sqrt{1-v^2}}
         (1,0,0,\Omega),\\
v &= (\Omega-\omega)r\sin\theta\,e^{-\rho}.
\end{aligned}
\end{equation}
The Einstein field equations~(which imply local energy-momentum conservation through the contracted Bianchi identity) then reduce to 5 coupled PDEs which can be solved given boundary conditions (central density, rotation speed, and asymptotic flatness) and an EoS~\cite{Komatsu1989,Gourgoulhon2010}. The functional forms of the PDEs are provided in the supplemental material~(Eqs.~\eqref{eq:supp-auxiliary-definitions},~\eqref{eq:supp-pde},~\eqref{eq:supp-differential-operators}). The solutions can have integral representations delineated as follows.

Define the log-enthalpy $H$ and coordinate equatorial radius
$r_e$ by~\cite{Cook1994,RNSManual}:
\begin{equation}
H(r,\theta) = H(p(r,\theta))=\int_0^{p(r,\theta)}\frac{dp'}{\epsilon(p')+p'},\,
H(r_e,\pi/2)=0.
\end{equation}
The three metric potentials $u\in\{\rho,\gamma,\omega\}$
have Green-function representations on angular bases
$Y_{u,n}$~\cite{Komatsu1989,Cook1994,StergioulasFriedman1995}:
\begin{equation}
\begin{aligned}
\bar{u}(s,\mu)
&=A_u(s,\mu)\sum_n Y_{u,n}(\mu)\\
&\quad\times\int_0^1\!\!\int_0^1
K_{u,n}(s,s',\mu')S_u(s',\mu')\,ds'\,d\mu',\label{Green's}
\end{aligned}
\end{equation}
where $s=r/(r+r_e)$, $\mu=\cos\theta$, and
$Y_{u,n}$, $A_u$, $K_{u,n}$, and $S_u$ are the corresponding
metric spherical harmonics, prefactors, Green kernels, and nonlinear sources~\cite{Cook1994,RNSManual} (see supplemental material, Eqs.~\eqref{eq:supp-angular-basis-definitions},~\eqref{eq:supp-prefactors},~\eqref{eq:supp-green-kernels},~\eqref{eq:supp-source-terms}). 
The fourth potential \(\alpha\) and the quantities of interest (mass \(M\) and equatorial radius \(R_e\)) are obtained from~\cite{Komatsu1989,Cook1994,Gourgoulhon2010}
\begin{equation}
\begin{aligned}
\alpha(r,\mu)
 &=\frac{\gamma(r,1)-\rho(r,1)}{2}
   -\int_\mu^1 S_\alpha(r,\mu')\,d\mu',\\
M&=\int_{\mathcal V}
 \left(-T^t{}_t+\sum_{i=r,\theta,\phi}T^i{}_i\right)
 \sqrt{-g}\,dr\,d\theta\,d\phi,\\
R_e&=\frac{1}{2\pi}\int_0^{2\pi}
 \sqrt{g_{\phi\phi}(r_e,\pi/2)}\,d\phi
 =r_e e^{(\gamma_e-\rho_e)/2}.
\end{aligned}
\end{equation}
Here $S_\alpha$ is the right-hand side of the
first-order equation $\partial_\mu\alpha=S_\alpha$~(defined in the supplemental material, Eq.~\eqref{eq:supp-alpha-source}),
$\mathcal V$ is the stellar interior, and
$\sqrt{-g}=e^{2\alpha+\gamma}r^2\sin\theta$~\cite{Komatsu1989,Gourgoulhon2010}.
Evaluating the matter-dependent sources and stress--energy
tensor requires EoS interpolation to obtain $\epsilon(H)$
and $p(H)$ from the hydrostationary enthalpy profile~\cite{Cook1994,RNSManual}.  \textbf{Note that the prefactors and source terms in Eq.~\eqref{Green's} as functions of  $(r,\mu)$ are dependent on the functional forms of the solutions $\rho(r,\mu),\gamma(r,\mu), \omega(r,\mu), \alpha(r,\mu), \Omega$} (see supplemental material, Eq.~\eqref{eq:supp-source-terms}).

\begin{figure*}
    \centering
    \includegraphics[width=0.48\textwidth]{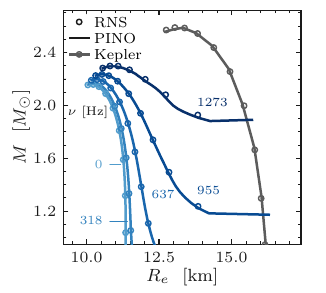}
    \includegraphics[width=0.48\textwidth]{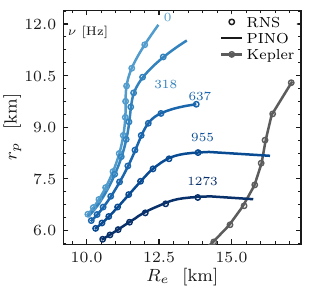}
    \caption{\label{fig:MRdemo}Demonstrating solver accuracy relative to \texttt{RNS}. In the \textit{left} panel, we show NS mass, and in the \textit{right} panel, polar radius, both as functions of equatorial radius and angular speed for the APR4\_EPP~\citep{Akmal1998, Read2009} EoS model. The ``Kepler'' limit denotes the angular speed above which the NS starts to lose mass at the equator due to insufficient centripetal acceleration.}
\end{figure*}

Numerical solvers~(such as the \texttt{RNS} code) start with a guess for \((\gamma,\rho,\omega)\), determine \(r_e\), \(\Omega\), and the matter profiles, compute \(\alpha\), and evaluate the Green-function
integrals~\cite{StergioulasFriedman1995,RNSManual}. The relaxed updates
$\mathbf{u\leftarrow c_f \bar{u}+(1-c_f)u}$ are repeated until
$r_e$ converges, after which $M$, $J$, and $R_e$
are evaluated~\cite{RNSManual}. 
Further details such as PDE and Green's function expressions are provided as supplemental material, and the numerical solutions~(obtained using the \texttt{RNS} code) are visualized in FIG.~\ref{fig:MRdemo}. \textbf{However, the relaxed update formalism leads to computationally costly numerical solvers}, which renders unbiased Monte Carlo EoS inference from uncertain $M-R_e-\Omega$ observations and high-fidelity spin evolution during mass transfer computationally prohibitive.

\emph{Physics-Informed Neural Operators} --- For accelerating PDE solutions, we propose using physics-informed neural operators~(PINOs)~\cite{Li2021PINO,Wang2021PIDeepONet}, which are universal approximators of a wide class of PDE solutions~\cite{Lu2021DeepONet}. We use the \texttt{DeepONet} architecture~\cite{Lu2021DeepONet} to model an operator that maps the EoS functional relationship (on a density grid) into metric-potential profiles (on a radius and latitude space):
\begin{equation}
\begin{aligned}
\begin{pmatrix}
\hat\gamma/r_e^2\\
\hat\rho/r_e^2\\
r_e\hat\omega
\end{pmatrix}
&=\vec d(s,q)\odot\Biggl[
\vec c +\vec a\odot\sum_{k=1}^{K}
\vec b_k(\vec z)\tanh t_k(s,\mu^2)
\Biggr],\\
\vec d(s,q)
&=\bigl(1-s,\,1-s,\,(1-s)^3\chi(q)\bigr)^{\mathsf T}.
\end{aligned}
\label{eq:DeepONet}
\end{equation}
Here $q=r_p/r_e$, the branch $\vec{b}$ and trunk $t$ are Multi-Layered Perceptrons~\citep[MLPs,][]{Goodfellow2016} with $_k$ denoting the kth output of the MLPs,  
and the branch input is
\[
\vec z=
\left(
\{\epsilon_j/\epsilon_c\}_{j=1}^{N^{\mathcal{E}}_{\rm grid}},
\{p_j/\epsilon_c\}_{j=1}^{N^{\mathcal{E}}_{\rm grid}},
h_c,q
\right),
\]
where $(\epsilon_j,p_j=p_{\mathcal{E}}(\epsilon)_j)$ are samples of the input EoS
$\mathcal E$. The branch network produces
$\vec b_k\in\mathbb R^3$ for each $k$, the trunk network produces
$t_k$, and $\vec a,\vec c\in\mathbb R^3$ are additional trainable
scales and offsets~(in addition to those of the MLPs); $\odot$ denotes componentwise
multiplication. We set $\chi(q)=1$ by default, or
$\chi(q)=\sqrt{\max(1-q,0)}$ when enforcing the
nonrotating limit. 
\begin{figure*}
    \centering
    \includegraphics[width=0.48\textwidth]{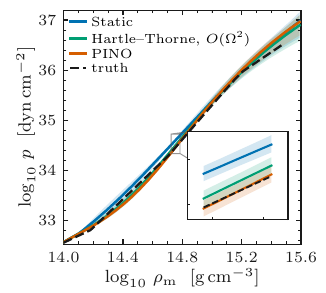}
    \includegraphics[width=0.48\textwidth]{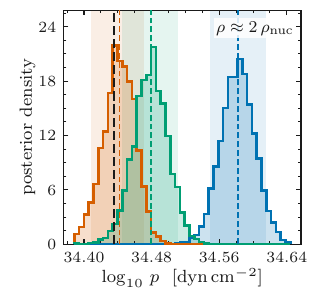}
    \caption{\label{fig:BHIMC} 
    Simulation-Recovery: EoS pressure vs. mass density~($\rho_m$) posteriors for the entire range of densities~(left) and at a fixed density~(twice the nuclear saturation density, right). \textbf{Ignoring rotation~(or using second-order corrections) excludes the ground truth in favour of stiffer EoSs at more than 90\% confidence in several regions where ours does not.}}
\end{figure*}

\textbf{Note that the \texttt{DeepONet} output replaces the relaxed update loop of RNS}, leading to mass-radius evaluations per EoS pressure-density curve that are faster by several orders of magnitude. To predict physical solutions, we train the \texttt{DeepONet}'s learnable parameters \textbf{in a fully unsupervised setting by minimizing a physics-only loss function}. Using the \texttt{DeepONet} output, we construct source terms and Green's functions of Eq.~\eqref{Green's} and obtain $\mathbf{\bar\gamma,\bar\rho,\bar\omega\leftarrow \hat\gamma,\hat\rho,\hat\omega}$. Then, the residuals

\begin{align}
  r_\gamma &= \frac{\bigl\langle (\bar\gamma-\hat\gamma)^2 \bigr\rangle}
                   {\tfrac12\langle\hat\gamma^2\rangle + \tfrac12\langle\bar\gamma^2\rangle + \epsilon},
\end{align}
\begin{align}
  r_\rho   &= \frac{\bigl\langle (\bar\rho-\hat\rho)^2 \bigr\rangle}
                   {\tfrac12\langle\hat\rho^2\rangle + \tfrac12\langle\bar\rho^2\rangle + \epsilon},
  \end{align}
\begin{align}
  r_\omega &= \frac{\bigl\langle (\bar\omega-\hat\omega)^2 \bigr\rangle}
                   {\tfrac12\langle\hat\omega^2\rangle + \tfrac12\langle\bar\omega^2\rangle
                    + 10^{-2}\,\hat\Omega^2 + \epsilon_\omega},
  \end{align}
\begin{align}
  \epsilon &= 10^{-4}, \epsilon_\omega = 10^{-6},
\end{align}
are used to compute the loss function:
\begin{equation}
  \mathcal{L}_{\rm res} = \frac{1}{N_{\rm sane}} \sum_{i\,\in\,\mathrm{sane}}
  \Bigl( r_\gamma^{(i)} + r_\rho^{(i)} + w_\omega\, r_\omega^{(i)} \Bigr),
  \qquad w_\omega = 1,
\end{equation}
where
\begin{equation}
  \langle q\rangle \equiv \frac{1}{N^s_{grid} N^\mu_{grid}}\sum_{j=1}^{N^s_{grid}}\sum_{m=1}^{N^\mu_{grid}} q(s_j,\mu_m).
\end{equation}

\begin{table}[t]
\centering
\caption{Relative errors, evaluation times, and speed-ups.}
\label{tab:solver_comparison}
\setlength{\tabcolsep}{3pt}
\renewcommand{\arraystretch}{1.15}
\resizebox{\columnwidth}{!}{%
\begin{tabular}{@{}lccccc@{}}
\hline\hline
Solver & $|\Delta M|$ [\%] & $|\Delta R|$ [\%]
& \multicolumn{2}{c}{Time} & Speed-up \\
& & & PINO & Reference & \\
\hline
\shortstack[l]{Rotating}
& $0.14^{+0.29}_{-0.09}$
& $0.03^{+0.26}_{-0.02}$
& $35\,\mathrm{ms}$ & $45\,\mathrm{s}$ & $\mathbf{1283\times}$ \\
\shortstack[l]{Static}
& $0.10^{+0.55}_{-0.05}$
& $0.05^{+0.35}_{-0.04}$
& $15\,\mathrm{ms}$ & $102\,\mathrm{ms}$ & $\mathbf{7\times}$ \\
\hline\hline
\end{tabular}%
}
\end{table}
\noindent
\textbf{At vanishing loss, the PINO output is a fixed point of the integral representation of the PDEs}. By minimizing this loss function using the Adam optimizer~\cite{KingmaBa2015}, a cosine restart scheduler~\cite{LoshchilovHutter2017} with amplitude decay, and early stopping~\cite{Prechelt1998EarlyStopping} on a validation loss plateau, we find optimal model parameters $(\vec{a}_k,\vec{b_k},$ and the weights and biases of the MLP$)$. \textbf{We note that, once trained, the \texttt{DeepONet} outputs are continuous functions of $r,\theta$ that can be evaluated at locations not restricted to the grid used in training~\cite{Lu2021DeepONet,Kovachki2023NO}.} Our training data comprises 40000 stars, each with its own EoS generated from the spectral parametrization~\cite{Lindblom2010Spectral} with physically meaningful priors (that enforce causality of the sound speed and thermodynamic stability, \citep{Abbott2018EOS}) and a central density randomly selected over the allowed range (denser than a white dwarf, rarer than the central density corresponding to the maximum mass) and a rotation speed randomly selected between the static and mass-shedding limits of said EoS. Further details about training, hyperparameter tuning, and the exact expressions of the branch and trunk MLPs are provided in the supplemental material~(Algorithm~\ref{alg:pino-training}, Eqs.~\eqref{eq:supp-mlp-input-layer}~-~\eqref{eq:supp-mlp-parameter-dimensions}).


Once trained, the solver can accurately predict $M-R_e$ curves given any EoS table (that is in distribution relative to the training set) and rotation period with a several-orders-of-magnitude speed-up, as demonstrated in FIG.~\ref{fig:MRdemo} and Table~\ref{tab:solver_comparison}. We further provide a much simpler PINO for the TOV equations is as part of our framework which is presented in the supplemental material~(Algorithm~\ref{alg:pino-training-tov}); it achieves a significant speed-up relative to numerical solvers, though smaller than in the axisymmetric case. While tabulated equilibria and supervised surrogates rely on numerical stellar models~\cite{Lavagetto2004,Fortin2016,Liu2026,Papigkiotis2025}, \textbf{our PINOs are trained once across the EoS prior without solution labels, mapping $p(\epsilon)$ and stellar parameters to continuous (mesh-free) stellar profiles evaluable on arbitrary coordinate grids}. 


\emph{Bayesian Hierarchical EoS Inference from Spin-mass-radius Observations} --- With our trained solvers, we can rapidly explore EoS models given data from multiple independent NS observations that facilitate spin and uncertain mass-radius measurements~\cite{LandryEssick2019,Essick2020,Landry2020,Raaijmakers2021}. The likelihood of an EoS model ($\mathcal{E}$) given independent and identically distributed observations is therefore~\cite{LackeyWade2015,Lange2018RIFT,Ghosh2021,GolombTalbot2022,Ray2023,Wysocki2020, Wagner2025RIFT, OShaughnessyZhu2026}:
\begin{equation}
    \mathcal{L}(\mathcal{E};\vec{d},\vec{\Omega}) \propto \prod_{i=1}^N\int_{M = \hat{f}(R_e,\Omega_i)}dR_e \frac{p(R_e, M|d_i)}{\pi_i(R_e|M)},
\end{equation}
where $p(R_e, M|d_i)$ is the posterior distribution of masses and equatorial radii for the $i$th observation (obtained using kernel density estimation from publicly released samples) and $\pi_i$ is the corresponding EoS-agnostic prior on the equatorial radius used during posterior estimation. 

Our fast solvers facilitate empirical constraints on the pressure-density plane using parametrized EoS models and sampling the posterior distribution of the EoS parameters obtained with physically motivated priors:
\begin{equation}
    p(\vec{\gamma}|\vec{d},\vec{\Omega}) \propto \pi(\vec{\gamma})\mathcal{L}(\mathcal{E}_{\vec{\gamma}};\vec{d},\vec{\Omega}),
\end{equation}
using Markov chain Monte Carlo~(MCMC). 

To benchmark the performance of our solvers in mitigating EoS inference biases in a controlled setting, we perform an injection-recovery analysis~(FIG.~\ref{fig:BHIMC}) using simulated mass-radius and spin measurements for ten NSs. We generate 30 injected configurations using independent \texttt{RNS} solutions for the APR4\_EPP~\citep{Akmal1998, Read2009} EoS, with spins drawn uniformly between $300$ and $1000\,\mathrm{Hz}$ and masses drawn from a truncated Galactic double-Gaussian distribution~\cite{Alsing2018}. We model the mass-radius measurements as uncorrelated Gaussians centered on the injected values, with $5\%$ fractional uncertainties in both mass and radius, adopting an optimistic scenario motivated by the anticipated sample size and measurement precision of future probes~\cite{Li2025eXTP}. The supplemental material describes both datasets~(Table~\ref{tab:real-data}) and also presents the EoS analysis of real events~(FIG.~\ref{fig:BHI2}). 
We note here that \textbf{numerical solvers render MCMC intractable, necessitating the use of PINO solvers for EoS inference that achieves the same fidelity}. 


\begin{figure}
    \centering
    \includegraphics[width=0.48\textwidth]{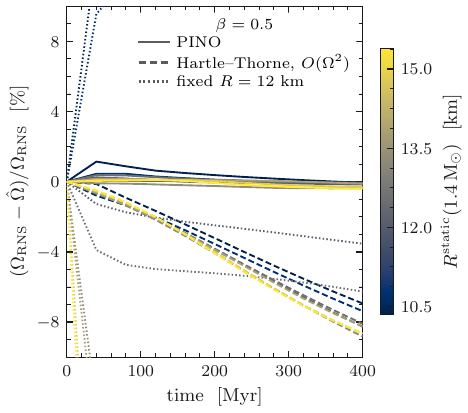}
    \caption{Impact of rotation-informed structure on accretion-driven spin-up for a toy mass-transfer model that assumes $\beta=0.5,\,\dot{M}_{tr}=10^{-9}\rm{M}_{\odot}\rm{yr}^{-1}$. Three treatments of structure calculations (fixed radius~\cite{Chattopadhyay2026}, second-order Hartle-Thorne~\cite{Holgado2021}, and PINO) coupled with accretion evolution are compared against the full \texttt{RNS}~\cite{StergioulasFriedman1995} coupled evolution. \textbf{PINO provides the smallest relative error at negligible computational cost.}}
    \label{fig:acc}
\end{figure}
\emph{Impact on Binary Evolution} --- On a population level, the accuracy and scalability of NS structure solvers impact several investigations in addition to EoS inference from mass-radius-spin measurements. In particular, NS structural response during accretion is expected to substantially affect the observable population properties of MSPs in binary systems~\cite{Possenti1998,Burderi1999,Possenti1999,Lavagetto2004,Bejger2011,Fortin2016,Holgado2021}. Accretion-driven spin-up during mass transfer is a key evolutionary channel that connects unknown population properties at the zero-age main sequence and NS birth to the observed ensembles in radio surveys~\cite{Kiel2008,Oslowski2011,Chattopadhyay2020,Sgalletta2023,Chattopadhyay2026,Sautron2026}. Modeling these channels is crucial for constraining binary evolution and natal distributions from large survey datasets~\cite{VignaGomez2018,Chattopadhyay2020,Sgalletta2023}; existing treatments range from fixed-radius prescriptions and approximate structural corrections to precomputed fully relativistic rotating-star equilibria~\cite{Chattopadhyay2021,Possenti1998,Possenti1999,Holgado2021,Lavagetto2004,Fortin2016}.

While tracking the EoS-informed radius in the spherical limit can substantially alter spin-up~\citep{Holgado2021} and hence the observed population of spun-up MSPs relative to treatments that fix the radius throughout mass transfer, \textbf{ignoring the effects of rotation or relying on approximate rotational corrections introduces additional biases that can accumulate over time}. Precomputed fully relativistic equilibria, while accurate, are costly to extend across a broad EoS prior, requiring dense grids.

For example, rotating structure can affect the spin-up prescriptions such as that in Ref.~\cite{Chattopadhyay2026}:
\begin{equation}
\begin{gathered}
\begin{pmatrix}
\dot{M}_b\\ \dot{J}\\ \dot{B}
\end{pmatrix}
=
\begin{pmatrix}
\beta(t)\dot M_{\rm tr}(t)\\[5pt]
\epsilon\dot M_b
\left(\sqrt{\frac{GMR_A}{2}}-\frac{\Omega R_A^2}{4}\right)
-\frac{8\pi B^2R_{\rm eq}^6\Omega^3\sin^2\alpha}{3\mu_0c^3}
\\[5pt]
-\frac{\dot M_b}{\kappa}(B-B_{\min})
\end{pmatrix},
\\[6pt]
(M,M_b, R_{\rm eq},\Omega,I)
=\mathcal S_{\rm EoS}(\varepsilon_c,J;\vec{\gamma}),
\end{gathered}
\end{equation}
where $M_b$ is the baryonic mass, $J$ is the stellar angular momentum, $B$ is the equatorial surface dipole magnetic-field strength, $R_{\rm eq}$ is the equatorial radius, and $I$ is the moment of inertia; $\dot M_{\rm tr}$ is the mass-transfer rate, $\beta$ is the accreted mass fraction, $\epsilon$ is the angular-momentum transfer efficiency, $R_A$ is the Alfv\'en radius, and $\alpha$ is the magnetic inclination angle. The function $\mathcal S_{\rm EoS}$ is the stellar-structure solver~(compared between PINO, static/TOV, and Hartle-Thorne~($O(\Omega^2)$ corrections to static solutions~\citep{Hartle1967, HartleThorne1968})). Ideally, $\beta(t)$ and $\dot{M}_{tr}(t)$ are supplied by detailed binary evolution calculations with an adopted accretion prescription~\citep{Fragos2023POSYDON,Andrews2025POSYDON,Chattaraj2026POSYDON}.

We demonstrate in FIG.~\ref{fig:acc} that \textbf{biases induced by fixed or approximate structure response accumulate over time} in the predicted accretion-driven spin-up with a toy mass-transfer model that assumes $\beta=0.5,\,\dot{M}_{tr}=10^{-9}\rm{M}_{\odot}\rm{yr}^{-1}$~\citep{Kar2024, Bejger2011}. \textbf{Repeated \texttt{RNS} solutions, even for a fixed EoS, can take hours per star, which is prohibitive for population synthesis}, and marginalizing EoS uncertainties over any prior or an independent posterior is intractable;\textbf{ our PINO solvers are necessary for high-fidelity spin predictions of observable MSP populations.}

\emph{Conclusion} --- In this work, we developed PINO solvers for axisymmetric general relativistic hydrostatic equilibrium, which achieve comparable accuracy to numerical solvers while being $\sim 10^3\times$ faster. Our solvers,\textbf{ trained once across the EoS prior solely by a physics-informed loss function, provide mesh-free stellar profiles with explicit $p(\epsilon)$ dependence}, complementing tabulated equilibria and supervised surrogates~\cite{Lavagetto2004,Fortin2016,Liu2026,Papigkiotis2025} and facilitating EoS inference and accretion-driven spin evolution modeling at scale with self-consistent rotation. They enable higher-fidelity empirical constraints on phenomenological EoS parameterizations and pulsar spin predictions during mass transfer, relative to static treatments and second-order rotational corrections.

While rotation-induced changes in MCMC constraints are modest relative to current measurement uncertainty, future observations will facilitate stronger constraints, with rotation-induced biases likely becoming non-negligible. In particular, the approved eXTP mission is expected to furnish tens of new observations of both rotation- and accretion-powered MSPs, providing highly precise radius measurements~(within 5--10\%) for a wide range of spins~\citep{Li2025eXTP}, with proposed missions~\citep{Cruise2025NewAthena, ColibriBursts, Ray2024STROBEX} promising expanded datasets. Alongside these advances, scalable structure calculations will enable population synthesis to self-consistently connect accretion-driven spin evolution to pulsar populations characterized by current~(FAST and MeerKAT~\citep{Han2021GPPS,Bailes2020MeerTime}) and upcoming~(SKA~\citep{Keane2015SKA}) large radio-timing surveys. Looking forward, \textbf{our framework presents a significant step towards scaling high-fidelity probes of degenerate nuclear matter and binary evolution to large datasets of rotating NS observations.}

\emph{Future Work} --- Extending the PINOs to differential rotation and magnetic fields will substantially broaden their scope; constraints on parametrized interior current~\cite{Pili2015,Pili2017} and angular momentum distributions~\cite{Uryu2017} can become computationally tractable. Magnetic equilibria linking interior currents to exterior fields and stellar deformations will enable constraints on parametrized magnetic configurations from magnetar X-ray spectra and polarization with emission modeling, and from continuous GWs if the magnetic deformation is misaligned with the spin~\cite{Pili2015,Taverna2022,FriebenRezzolla2012,Pili2017,Uryu2017}. Differentially rotating equilibria would support modeling of merger-remnant collapse thresholds and, coupled with finite-temperature and oscillation-mode treatments, the interpretation of postmerger GWs anticipated with next-generation detectors~\cite{Uryu2017,BausweinStergioulas2017,Breschi2025}.

Parallel investigations with the current solvers include ongoing efforts to incorporate GW measurements~\cite{Dietrich2020,Landry2020,Raaijmakers2021} and marginalize over population-informed priors~\cite{GolombTalbot2022, Wysocki2020, Raaijmakers2021}. We aim to expand the training set with non-parametric pressure-density priors~\cite{Landry2020,Legred2021} and hybrid models allowing quark-matter phase transitions~\cite{Legred2021}, and plan end-to-end simulation-recovery studies to explore systematics. Finally, ongoing efforts to incorporate the stellar-structure response to accretion~\cite{Lavagetto2004,Fortin2016,Holgado2021} into detailed binary evolution frameworks such as \texttt{POSYDON}~\cite{Fragos2023POSYDON,Andrews2025POSYDON, Chattaraj2026POSYDON} will achieve high-fidelity modeling of spin evolution during mass transfer, connecting initial stellar distributions, NS properties at birth, binary evolution physics, and dense-matter properties to observations~\cite{Possenti1999,Chattopadhyay2020,Sgalletta2023,Chattopadhyay2026}.

\emph{Code Availability and Software} --- The code developed to implement the solvers and all associated analyses will be made publicly available at \url{https://github.com/AnaryaRay1/pinoTOV} upon acceptance and, in the meantime, upon reasonable request. It relies on the \texttt{numpy}~\cite{Harris2020NumPy}, \texttt{scipy}~\cite{Virtanen2020SciPy}, \texttt{pytorch}~\cite{Paszke2019PyTorch}, and \texttt{matplotlib}~\cite{Hunter2007Matplotlib} libraries. We have used the \texttt{LALSimulation} library~\citep[][\url{https://git.ligo.org/lscsoft/lalsuite}]{LALSuite2018} to generate EoS tables and provide the static~(ToV) benchmark, the \texttt{RNS} library~\citep[][\url{https://github.com/cgca/rns.git}]{StergioulasFriedman1995} as the numerical solver benchmark, and the \texttt{QLIMR} library~\citep[][\url{https://gitlab.com/nsf-muses/ns-qlimr/qlimr}]{QLIMR} for the Hartle-Thorne~\citep{Hartle1967,HartleThorne1968} benchmark.

\emph{Acknowledgement} --- We thank Ish Gupta, Phillip Srivastava, Seth Gosage, and Jolien Creighton for useful discussions and suggestions. This work was supported by the National Science Foundation~(NSF) award PHY-2207945. We are grateful for the computational resources provided by the LIGO laboratory and supported by National Science Foundation Grants PHY-0757058 and PHY-0823459. We gratefully acknowledge the support of the NSF-Simons AI-Institute for the Sky (SkAI) via grants NSF AST-2421845 and Simons Foundation MPS-AI-00010513.

\emph{Disclaimer} --- We acknowledge the use of Large Language Models~(LLMs) in this work, specifically Anthropic's Claude (Fable-5.1) for debugging code, strategizing optimal implementations, and improving scalability, and OpenAI's ChatGPT~(Astra 6.1) for generating plotting code, polishing text, reviewing literature, and generating Overleaf commands. LLMs were not used for any other aspect of this manuscript. The author claims sole responsibility for idea conception, the manuscript text, and all associated code, results, and claims. Every line of LLM-generated code and text, every citation, and every Overleaf command has been checked for correctness and potential plagiarism by the author to the best of their knowledge.

\nocite{*}

\bibliography{apssamp}
\appendix
\section{Supplementl Material}

\noindent
Here we provide additional details and results for our framework and analysis.

\noindent
\textbf{PDEs and Green's functions.} $\,$\paragraph{PDEs and Green's functions.}
We use geometrized units, $G=c=1$, and the metric,
perfect-fluid stress--energy tensor, log-enthalpy, and
coordinate equatorial radius defined in the main text,
which we repeat here for the sake of continuity:
\begin{equation}
\begin{aligned}
ds^2 &= -e^{\gamma+\rho}dt^2
       +e^{2\alpha}(dr^2+r^2d\theta^2)\\
     &\quad +e^{\gamma-\rho}r^2\sin^2\theta
       (d\phi-\omega\,dt)^2,\\
T^{\mu\nu} &= (\epsilon+p)u^\mu u^\nu+p\,g^{\mu\nu},\\
u^\mu &= \frac{e^{-(\gamma+\rho)/2}}{\sqrt{1-v^2}}
         (1,0,0,\Omega),\\
v &= (\Omega-\omega)r\sin\theta\,e^{-\rho}.
\end{aligned}
\label{eq:supp-metric-matter}
\end{equation}
\begin{equation}
\begin{aligned}
H(r,\theta)
 &=H(p(r,\theta))
   =\int_0^{p(r,\theta)}
     \frac{dp'}{\epsilon(p')+p'},\\
H(r_e,\pi/2)&=0.
\end{aligned}
\label{eq:supp-enthalpy-surface}
\end{equation}
The potentials $\gamma,\rho,\alpha,\omega$ in
Eq.~\eqref{eq:supp-metric-matter} are physical
metric functions; in particular, $\rho$ is not the matter
density. Define
\begin{equation}
\begin{aligned}
\nu&=\frac{\gamma+\rho}{2},
& Q&=\frac{\epsilon+p}{1-v^2},\\
v&=(\Omega-\omega)r\sin\theta\,e^{-\rho},
& W&=r^2\sin^2\theta\,e^{-2\rho}(\partial\omega)^2.
\end{aligned}
\label{eq:supp-auxiliary-definitions}
\end{equation}
Here $\epsilon$ and $p$ are the fluid-frame energy density
and pressure, and $\Omega$ is the uniform angular velocity
relative to infinity.

The equilibrium equations can be written as
\cite{Gourgoulhon2010}
\begin{equation}
\begin{aligned}
\Delta_3\nu
 &=4\pi e^{2\alpha}[Q(1+v^2)+2p]
   +\tfrac12 W-\partial\nu\,\partial\gamma,\\
\mathcal D_4 e^\gamma
 &=16\pi e^{2\alpha+\gamma}p,\\
\mathcal D_5\omega
 &=-16\pi e^{2\alpha}Q(\Omega-\omega)
   +\partial\omega\,\partial(4\nu-3\gamma),\\
\Delta_2(\alpha+\nu)
 &=8\pi e^{2\alpha}(p+Qv^2)
   +\tfrac34 W-(\partial\nu)^2,\\
\frac{\partial_i p}{\epsilon+p}
 &=-\partial_i\!\left[\nu+\tfrac12\ln(1-v^2)\right],
\end{aligned}
\label{eq:supp-pde}
\end{equation}
closed by $\epsilon=\epsilon(p)$ and
$\Omega=\mathrm{const}$. The fifth equation is a
first-order hydrostationary equation, with
$i\in\{r,\theta\}$.

The differential operators are
\begin{equation}
\begin{aligned}
\Delta_2
 &=\partial_r^2+\frac1r\partial_r
   +\frac1{r^2}\partial_\theta^2,\\
\Delta_3
 &=\partial_r^2+\frac2r\partial_r
   +\frac1{r^2}
      (\partial_\theta^2+\cot\theta\,\partial_\theta),\\
\mathcal T
 &=\frac1r\partial_r
   +\frac{\cot\theta}{r^2}\partial_\theta,\\
\mathcal D_d&=\Delta_3+(d-3)\mathcal T,\\
\partial f\,\partial g
 &=f_{,r}g_{,r}+\frac1{r^2}f_{,\theta}g_{,\theta},
\qquad
(\partial f)^2=\partial f\,\partial f .
\end{aligned}
\label{eq:supp-differential-operators}
\end{equation}
We use $\mathcal D_3=\Delta_3$ below.

\paragraph{Matter and surface conditions.}
Integration of the hydrostationary equation gives
\cite{Cook1994,Gourgoulhon2010}
\begin{equation}
\begin{aligned}
H(r,\theta)
 &=C-\nu(r,\theta)-\tfrac12\ln(1-v^2),\\
C&=H_c+\nu_c=\nu_p.
\end{aligned}
\label{eq:supp-hydrostationary-integral}
\end{equation}
The subscripts $c,p,e$ denote the centre, polar surface,
and equatorial surface. The coordinate surface radii obey,
consistently with Eq.~\eqref{eq:supp-enthalpy-surface},
\begin{equation}
H(r_p,0)=0,\qquad H(r_e,\pi/2)=0.
\label{eq:supp-surface-radii}
\end{equation}
The stellar interior $\mathcal V$ is the connected
matter region containing the centre. Within $\mathcal V$,
EoS interpolation supplies $\epsilon(H)$ and $p(H)$;
outside it, $\epsilon=p=0$. The Bernoulli expression
continued outside the star need not represent a physical
fluid enthalpy.

At the centre and pole $v=0$, while the equatorial
surface condition yields
\begin{equation}
\begin{aligned}
2H_c&=(\gamma+\rho)_p-(\gamma+\rho)_c,\\
\Omega
 &=\omega_e+\frac{e^{\rho_e}}{r_e}
   \sqrt{1-e^{2(\nu_p-\nu_e)}} .
\end{aligned}
\label{eq:supp-centre-pole-rotation}
\end{equation}
The positive root selects the corresponding orientation
of rotation. These are equilibrium relations: the
coordinate radius $r_e$ is determined jointly with the
fields and the surface, rather than independently by
the centre--pole relation.

\paragraph{Elliptic equations used for the Green map.}
The first three field equations can be rearranged into
the KEH form~\cite{Komatsu1989,Cook1994}
\begin{equation}
\begin{aligned}
\mathcal D_3(\rho e^{\gamma/2})&=S_\rho,\\
\mathcal D_4(\gamma e^{\gamma/2})&=S_\gamma,\\
\mathcal D_5(\omega e^{\gamma/2-\rho})&=S_\omega.
\end{aligned}
\label{eq:supp-keh-equations}
\end{equation}
With the operator conventions above, the sources are
\begin{equation}
\begin{aligned}
S_\rho={}&e^{\gamma/2}
 \biggl\{8\pi e^{2\alpha}Q(1+v^2)+W+\mathcal T\gamma\\
 &\quad+\rho\left[
 8\pi e^{2\alpha}p-\tfrac12\mathcal T\gamma
 -\tfrac14(\partial\gamma)^2\right]\biggr\},\\
S_\gamma={}&e^{\gamma/2}
 \left[16\pi e^{2\alpha}p(1+\gamma/2)
       -\tfrac{\gamma}{4}(\partial\gamma)^2\right],\\
S_\omega={}&e^{\gamma/2-\rho}
 \biggl\{-16\pi e^{2\alpha}Q\Omega\\
 &\quad+\omega\bigl[
 8\pi e^{2\alpha}(\epsilon+2p)-W
 +(\partial\rho)^2-\tfrac14(\partial\gamma)^2\\
 &\hspace{38mm}
 -2\mathcal T\rho-\tfrac12\mathcal T\gamma
 \bigr]\biggr\}.
\end{aligned}
\label{eq:supp-source-terms}
\end{equation}
All quantities on these right-hand sides are evaluated
from the input fields and their associated matter profiles.

\paragraph{Angular bases, prefactors, and kernels.}
Assuming equatorial reflection symmetry, let
\begin{equation}
\begin{aligned}
d_\rho&=3,\qquad d_\gamma=4,\qquad d_\omega=5,\\
Y_{u,n}(\mu)
 &=C_{2n}^{(d_u-2)/2}(\mu),
 \qquad n=0,1,\ldots,\\
w_d(\mu)&=(1-\mu^2)^{(d-3)/2},\\
N_{d,n}
 &=\int_0^1 w_d(\mu)
       [C_{2n}^{(d-2)/2}(\mu)]^2\,d\mu .
\end{aligned}
\label{eq:supp-angular-basis-definitions}
\end{equation}
Here $C_\ell^\lambda$ is a Gegenbauer polynomial. Explicitly,
\begin{equation}
\begin{aligned}
Y_{\rho,n}(\mu)&=P_{2n}(\mu),\\
Y_{\gamma,n}(\cos\theta)
 &=\frac{\sin[(2n+1)\theta]}{\sin\theta},\\
Y_{\omega,n}(\mu)&=\frac{dP_{2n+1}(\mu)}{d\mu},\\
N_{3,n}&=\frac1{4n+1},\qquad N_{4,n}=\frac{\pi}{4},\\
N_{5,n}&=\frac{(2n+1)(2n+2)}{4n+3}.
\end{aligned}
\label{eq:supp-angular-bases}
\end{equation}
Values at the axis are understood by continuity.

The metric prefactors in Eq.~\eqref{eq:supp-green} are
\begin{equation}
A_\rho=A_\gamma=e^{-\gamma/2},
\qquad
A_\omega=e^{\rho-\gamma/2}.
\label{eq:supp-prefactors}
\end{equation}
The minus sign from inversion of the elliptic operators
is included in the kernels below.

Define
\begin{equation}
\begin{aligned}
r(s)&=\frac{r_es}{1-s},
& \frac{dr'}{ds'}&=\frac{r_e}{(1-s')^2},\\
r_<&=\min\{r(s),r(s')\},
& r_>&=\max\{r(s),r(s')\}.
\end{aligned}
\label{eq:supp-radial-coordinates}
\end{equation}
Then the kernels corresponding to the physical-coordinate
sources defined above are
\begin{equation}
\begin{aligned}
K_{u,n}(s,s',\mu')
={}&-\frac{w_{d_u}(\mu')Y_{u,n}(\mu')}
 {(4n+d_u-2)N_{d_u,n}}\\
&\times
 \frac{r_<^{2n}}{r_>^{2n+d_u-2}}\,
 [r(s')]^{d_u-1}\frac{r_e}{(1-s')^2}.
\end{aligned}
\label{eq:supp-green-kernels}
\end{equation}
Thus
\begin{equation}
\begin{aligned}
\bar u(s,\mu)
={}&A_u(s,\mu)\sum_{n=0}^{\infty}Y_{u,n}(\mu)\\
&\times\int_0^1\!\!\int_0^1
 K_{u,n}(s,s',\mu')S_u(s',\mu')\,ds'\,d\mu'.
\end{aligned}
\label{eq:supp-green}
\end{equation}
Here $S_u(s',\mu')$ means the physical-coordinate source
evaluated at $r'=r_es'/(1-s')$ and
$\theta'=\arccos\mu'$. The powers of $r'$ and the
compactification Jacobian are already included in $K_{u,n}$
and must not be included a second time in $S_u$.
This is a normalization of the usual KEH expansions
\cite{Komatsu1989,Cook1994,StergioulasFriedman1995}.

These kernels impose regularity at the centre and decay
at infinity. The integration covers the entire spatial
domain, since the nonlinear gravitational sources need
not vanish outside the matter. Numerical implementations
truncate the angular sum and discretize the integrals.

\paragraph{First-order equation for $\alpha$.}
RNS obtains $\alpha$ from a first-order combination of
the Einstein equations rather than separately inverting
the elliptic equation for $\alpha+\nu$
\cite{Komatsu1989,RNSManual}. For a compact expression,
define
\begin{equation}
\begin{aligned}
D&=r\partial_r=s(1-s)\partial_s,\qquad m=1-\mu^2,\\
a&=1+D\gamma,\qquad b=-\mu+m\gamma_{,\mu},\\
\mathcal B&=ma^2+b^2.
\end{aligned}
\label{eq:supp-alpha-auxiliary-definitions}
\end{equation}
The symbol $m$ here is only an angular abbreviation,
not a mass. Introduce
\begin{equation}
\begin{aligned}
F={}&\tfrac12\bigl[
 D^2\gamma+(D\gamma)^2-m\gamma_{,\mu\mu}
 +2\mu\gamma_{,\mu}-b\gamma_{,\mu}\bigr]\\
 &+(D\nu)^2-m\nu_{,\mu}^2\\
 &-\tfrac14mr^2e^{-2\rho}
       \bigl[(D\omega)^2-m\omega_{,\mu}^2\bigr],\\
G={}&-m\bigl[
 2(D\nu)\nu_{,\mu}+D(\gamma_{,\mu})
 +(D\gamma)\gamma_{,\mu}\bigr]\\
 &+\mu D\gamma
 +\tfrac12m^2r^2e^{-2\rho}(D\omega)\omega_{,\mu}.
\end{aligned}
\label{eq:supp-alpha-source-components}
\end{equation}
Then
\begin{equation}
S_\alpha=-\nu_{,\mu}-\frac{bF+aG}{\mathcal B},
\qquad
\partial_\mu\alpha=S_\alpha.
\label{eq:supp-alpha-source}
\end{equation}
Axis regularity fixes the integration constant:
\begin{equation}
\begin{aligned}
\alpha(r,1)&=\frac{\gamma(r,1)-\rho(r,1)}{2},\\
\alpha(r,\mu)
 &=\frac{\gamma(r,1)-\rho(r,1)}{2}
   -\int_\mu^1 S_\alpha(r,\mu')\,d\mu'.
\end{aligned}
\label{eq:supp-alpha-integral}
\end{equation}
At the centre, the expression is interpreted through
its regular limit. For spherical configurations,
$S_\alpha=0$ and $\alpha=(\gamma-\rho)/2$.
The converged equilibrium also satisfies the elliptic
equation for $\alpha+\nu$ given above.

\paragraph{Self-consistency and observables.}
The bar in $\bar u$ denotes an integral-equation output,
not a rescaling or an independent physical potential.
A self-consistent solution obeys $\bar u=u$, the
hydrostationary equation, the EoS, and the surface conditions.
RNS approaches this solution through relaxed updates,
whereas a physics-informed model can enforce the same
conditions through residuals.

For completeness, the main-text mass integrand is
\begin{equation}
\begin{aligned}
-T^t{}_t+\sum_{i=r,\theta,\phi}T^i{}_i
={}&Q(1+v^2)+2p\\
 &+2\omega r\sin\theta\,e^{-\rho}Qv,
\end{aligned}
\label{eq:supp-mass-integrand}
\end{equation}
and
\begin{equation}
\sqrt{-g}=e^{2\alpha+\gamma}r^2\sin\theta,
\qquad
R_e=r_e e^{(\gamma_e-\rho_e)/2}.
\label{eq:supp-volume-radius}
\end{equation}
Here $r_e$ is the coordinate equatorial surface radius
defined in Eq.~\eqref{eq:supp-enthalpy-surface},
while $R_e$ is its circumferential radius. All observables
are evaluated from the self-consistent physical fields
\cite{Gourgoulhon2010}.

\noindent
\textbf{DeepONet Architecture, Algorithm and Training} --- The architecture, training, and tuning of our \texttt{DeepOnet} solver is summarized as follows.

\paragraph{Architecture.}
We use the \texttt{DeepONet} architecture defined in the
main text, which we repeat here for the sake of continuity:
\begin{equation}
\begin{aligned}
\begin{pmatrix}
\hat\gamma/r_e^2\\
\hat\rho/r_e^2\\
r_e\hat\omega
\end{pmatrix}
&=\vec d(s,q)\odot\Biggl[
\vec c+\vec a\odot\sum_{k=1}^{K}
\vec b_k(\vec z)\tanh t_k(s,\mu^2)
\Biggr],\\
\vec d(s,q)
&=\bigl(1-s,\,1-s,\,(1-s)^3\chi(q)\bigr)^{\mathsf T}.
\end{aligned}
\label{eq:supp-deeponet}
\end{equation}
Here $q=r_p/r_e$, and the branch and trunk are
multilayer perceptrons~\citep[MLPs,][]{Goodfellow2016}.
The branch input is
\begin{equation}
\vec z=
\left(
\{\epsilon_j/\epsilon_c\}_{j=1}^{N^{\mathcal E}_{\rm grid}},
\{p_j/\epsilon_c\}_{j=1}^{N^{\mathcal E}_{\rm grid}},
h_c,q
\right),
\label{eq:supp-branch-input}
\end{equation}
where $(\epsilon_j,p_j=p_{\mathcal E}(\epsilon_j))$ are
samples of the input EoS $\mathcal E$,
$\epsilon_c$ is the central energy density, and
$h_c=H_c$ denotes the central log-enthalpy.
The branch network produces $\vec b_k\in\mathbb R^3$
for each $k$, the trunk network produces $t_k$, and
$\vec a,\vec c\in\mathbb R^3$ are additional trainable
scales and offsets; $\odot$ denotes componentwise
multiplication. We set
\begin{equation}
\chi(q)=
\begin{cases}
1, & \text{by default},\\
\sqrt{\max(1-q,0)}, & \text{to enforce vanishing frame dragging at }q=1.
\end{cases}
\label{eq:supp-rotation-factor}
\end{equation}

\paragraph{MLP functional forms.}
The branch and trunk use the same MLP construction with
independent parameters. For either network
$\eta\in\{\mathrm b,\mathrm t\}$, let $w_\eta$ be its
hidden width and $L_\eta$ the number of hidden-to-hidden
affine layers. The input projection and activation are
\begin{equation}
\vec h_\eta^{(0)}(\vec x)
=\tanh\!\left(
W_\eta^{(0)}\vec x+\vec\beta_\eta^{(0)}
\right).
\label{eq:supp-mlp-input-layer}
\end{equation}
The remaining hidden layers are
\begin{equation}
\begin{aligned}
\vec g_\eta^{(\ell)}
&=\tanh\!\left(
W_\eta^{(\ell)}\vec h_\eta^{(\ell-1)}
+\vec\beta_\eta^{(\ell)}
\right),\\
\vec h_\eta^{(\ell)}
&=
\begin{cases}
\vec h_\eta^{(\ell-1)}+\vec g_\eta^{(\ell)},
 & \ell\ \text{odd},\\
\vec g_\eta^{(\ell)},
 & \ell\ \text{even},
\end{cases}\\
&\hspace{25mm}\ell=1,\ldots,L_\eta.
\end{aligned}
\label{eq:supp-mlp-hidden-layers}
\end{equation}
All activations act componentwise. This implements the
skip connections in \texttt{no.py}: an identity connection
is added at each odd-indexed hidden-to-hidden layer.
The output layer is affine:
\begin{equation}
\operatorname{MLP}_\eta(\vec x)
=W_\eta^{\rm out}\vec h_\eta^{(L_\eta)}(\vec x)
+\vec\beta_\eta^{\rm out}.
\label{eq:supp-mlp-output-layer}
\end{equation}
Thus the code argument \texttt{Nlayers} counts the
hidden-to-hidden layers, in addition to the initial
input projection; there are $L_\eta+1$ activated layers
and one affine output layer.

The input and output dimensions are
\begin{equation}
\begin{aligned}
d_{\mathrm b}^{\rm in}
 &=2N^{\mathcal E}_{\rm grid}+2,
& d_{\mathrm b}^{\rm out}&=3K,\\
d_{\mathrm t}^{\rm in}&=2,
& d_{\mathrm t}^{\rm out}&=K.
\end{aligned}
\label{eq:supp-mlp-dimensions}
\end{equation}
Accordingly, the parameter dimensions are
\begin{equation}
\begin{aligned}
W_\eta^{(0)}
 &\in\mathbb R^{w_\eta\times d_\eta^{\rm in}},
& \vec\beta_\eta^{(0)}
 &\in\mathbb R^{w_\eta},\\
W_\eta^{(\ell)}
 &\in\mathbb R^{w_\eta\times w_\eta},
& \vec\beta_\eta^{(\ell)}
 &\in\mathbb R^{w_\eta},\\
W_\eta^{\rm out}
 &\in\mathbb R^{d_\eta^{\rm out}\times w_\eta},
& \vec\beta_\eta^{\rm out}
 &\in\mathbb R^{d_\eta^{\rm out}}.
\end{aligned}
\label{eq:supp-mlp-parameter-dimensions}
\end{equation}
\begin{algorithm*}[t]
\caption{Physics-Informed DeepONet Training for Uniformly Rotating Axisymmetric Solutions}
\label{alg:pino-training}
\begin{algorithmic}[1]
\Require $\mathcal D=\{(\mathcal E_i,\epsilon_{c,i},q_i)\}_{i=1}^{40000}$:
EoS curves, central energy densities, and axis ratios
\Require $\mathcal D_{\rm train},\mathcal D_{\rm val}$:
training and validation subsets of $\mathcal D$
\Require $\mathcal A$: candidate DeepONet architectures
\Require $(\textsc{Adam},\eta_0,\beta_1,\beta_2)$:
optimizer and associated parameters
\Require $\textsc{CosineScheduler}$:
cosine restarts with amplitude decay
\Require $\mathcal X=\{(s_j,\mu_j)\}_{j=1}^{N_{\rm grid}}$:
collocation grid
\Require $\mathcal L_{\rm physics}$:
main-text residual loss
\Require $N_{\rm epoch},P$:
maximum epochs and early-stopping patience
\Ensure Trained $\texttt{DeepONet}^{\star}$

\State $L^\star_{\rm val}\gets+\infty$
\For{$A\in\mathcal A$}
    \State Initialize $\texttt{DeepONet}_A$ with parameters $\Theta$
    \State Initialize Adam and cosine scheduler
    \State $L_{A,\rm best}\gets+\infty$; $n_{\rm plateau}\gets0$
    \For{$e=1$ to $N_{\rm epoch}$}
        \For{minibatch $\mathcal B\subset\mathcal D_{\rm train}$}
            \State $L\gets\Call{PhysicsLoss}{\Theta,\mathcal B,\mathcal X}$
            \State $\Theta\gets
            \textsc{AdamStep}(\Theta,\nabla_\Theta L,\eta,\beta_1,\beta_2)$
            \State Update $\eta$ using $\textsc{CosineScheduler}$
        \EndFor
        \State $L_{\rm val}\gets
        \Call{PhysicsLoss}{\Theta,\mathcal D_{\rm val},\mathcal X}$
        \If{$L_{\rm val}<L_{A,\rm best}$}
            \State $L_{A,\rm best}\gets L_{\rm val}$;
            $n_{\rm plateau}\gets0$
        \Else
            \State $n_{\rm plateau}\gets n_{\rm plateau}+1$
        \EndIf
        \If{$L_{\rm val}<L^\star_{\rm val}$}
            \State $L^\star_{\rm val}\gets L_{\rm val}$
            \State $\texttt{DeepONet}^{\star}\gets
            \textsc{Copy}(\texttt{DeepONet}_A,\Theta)$
        \EndIf
        \If{$n_{\rm plateau}\geq P$}
            \State \textbf{break}
        \EndIf
    \EndFor
\EndFor
\State \Return $\texttt{DeepONet}^{\star}$

\Function{PhysicsLoss}{$\Theta,\mathcal B,\mathcal X$}
    \For{$(\mathcal E_i,\epsilon_{c,i},q_i)\in\mathcal B$}
        \State Construct $h_{c,i}$ and branch input $\vec z_i$
        \State Evaluate $\texttt{DeepONet}_{A,\Theta}$ on $\mathcal X$
        \State Determine $r_e,\Omega,H,\epsilon,p,\alpha$
        from the predicted fields and $\mathcal E_i$
        \State Evaluate $S_u,A_u,K_{u,n}$ for
        $u\in\{\rho,\gamma,\omega\}$
        \State Compute $\bar u$ using Eq.~\eqref{eq:supp-green}
        \State $\ell_i\gets\mathcal L_{\rm physics}$
        evaluated from the predictions and integral-equation outputs
    \EndFor
    \State \Return $|\mathcal B|^{-1}\sum_{i\in\mathcal B}\ell_i$
\EndFunction
\end{algorithmic}
\end{algorithm*}
\begin{figure*}
    \includegraphics[width=0.98\textwidth]{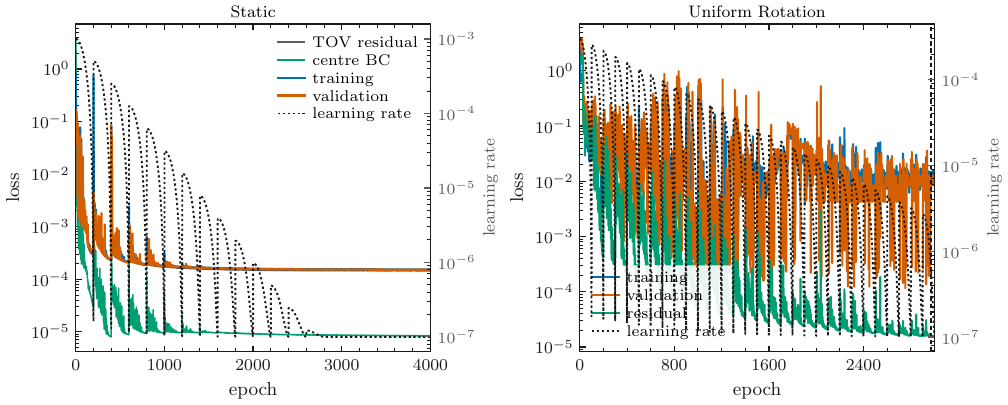}
    \caption{\label{fig:training}Loss and learning rate evolution during training for the Spherically symmetric~(\textit{left}) and uniformly rotating~\textit{(right)} solvers.}
\end{figure*}
\paragraph{Branch coefficients and trunk features.}
Write the branch output as three consecutive blocks:
\begin{equation}
\begin{aligned}
\operatorname{MLP}_{\mathrm b}(\vec z)
=\bigl(
 &b_{\gamma,1}(\vec z),\ldots,b_{\gamma,K}(\vec z),\\
 &b_{\rho,1}(\vec z),\ldots,b_{\rho,K}(\vec z),\\
 &b_{\omega,1}(\vec z),\ldots,b_{\omega,K}(\vec z)
\bigr)^{\mathsf T}.
\end{aligned}
\label{eq:supp-branch-output}
\end{equation}
The coefficient vectors in Eq.~\eqref{eq:supp-deeponet}
are therefore
\begin{equation}
\vec b_k(\vec z)=
\begin{pmatrix}
b_{\gamma,k}(\vec z)\\
b_{\rho,k}(\vec z)\\
b_{\omega,k}(\vec z)
\end{pmatrix},
\qquad k=1,\ldots,K.
\label{eq:supp-branch-coefficients}
\end{equation}
The trunk output is
\begin{equation}
\operatorname{MLP}_{\mathrm t}
\!\begin{pmatrix}s\\ \mu^2\end{pmatrix}
=
\begin{pmatrix}
t_1(s,\mu^2)\\
\vdots\\
t_K(s,\mu^2)
\end{pmatrix}.
\label{eq:supp-trunk-output}
\end{equation}
An additional $\tanh$ is applied to this affine output
before contraction with the branch coefficients:
\begin{equation}
\tau_k(s,\mu)=\tanh t_k(s,\mu^2).
\label{eq:supp-trunk-features}
\end{equation}
The same $K$ trunk features are shared by the three
potentials, while their branch coefficients, scales,
and offsets are distinct.

\paragraph{Architectural constraints.}
The dependence on $\mu^2$ enforces equatorial reflection
symmetry:
\begin{equation}
\hat u(s,-\mu)=\hat u(s,\mu),
\qquad
u\in\{\gamma,\rho,\omega\}.
\label{eq:supp-equatorial-symmetry}
\end{equation}
Since $1-s=r_e/(r+r_e)$, the radial prefactors in
Eq.~\eqref{eq:supp-deeponet} enforce
\begin{equation}
\hat\gamma,\hat\rho=O(r^{-1}),
\qquad
\hat\omega=O(r^{-3}),
\qquad r\rightarrow\infty,
\label{eq:supp-asymptotic-decay}
\end{equation}
for fixed stellar inputs and finite network parameters.
When the second choice in Eq.~\eqref{eq:supp-rotation-factor}
is used, $\hat\omega=0$ at $q=1$.

\emph{Training: Data and method} We generate $40{,}000$ stellar configurations by sampling spectral EoSs subject to causality and stability constraints, together with central energy densities and axis ratios. Given any architecture choice, we find optimal model parameters $(\vec{a}_k,\vec{b}_k,$ and the weights and biases of the MLP$)$ by minimizing the loss function using the Adam optimizer~\cite{KingmaBa2015}, a cosine restart scheduler~\cite{LoshchilovHutter2017} with amplitude decay, and early stopping~\cite{Prechelt1998EarlyStopping} on a validation loss plateau. We use an 80:20 training-validation split. We loop over architecture choices to select the solver with the lowest loss on a separately generated test set. The full workflow is summarized in Algorithm~\ref{alg:pino-training}, and loss evolution for the best architecture is visualized in FIG.~\ref{fig:training}.

\noindent
\textbf{Spherical Solvers} --- In geometrized units ($G=c=1$), spherical stellar equilibrium is governed by the Tolman--Oppenheimer--Volkoff equations~\cite{Tolman1939,OppenheimerVolkoff1939},
\begin{equation}
\frac{dp}{dr}
=-\frac{(\varepsilon+p)(m+4\pi r^3p)}{r(r-2m)},
\qquad
\frac{dm}{dr}=4\pi r^2\varepsilon,
\end{equation}
closed by the barotropic EoS,
\begin{equation}
\varepsilon=\varepsilon(p).
\end{equation}
Here $m(r)$ is the enclosed gravitational mass, with $m(0)=0$ and $p(R)=0$. Following the enthalpy formulation~\cite{Lindblom1992}, we introduce $h(p)=\int_0^p dp'/[\varepsilon(p')+p']$ and $x=h/h_c$, placing the surface at $x=0$ and the centre at $x=1$. We rescale each star using its central energy density: $\tilde\varepsilon=\varepsilon/\varepsilon_c$, $\tilde p=p/\varepsilon_c$, $\tilde r=r\sqrt{\varepsilon_c}$, and $\tilde m=m\sqrt{\varepsilon_c}$.

The stellar solution is reconstructed from a single positive field $g(x)=-d\tilde r^2/dx$ through
\begin{equation}
\begin{aligned}
\tilde r^2(x)&=\int_x^1 g(u)\,du,\\
\tilde m(x)&=-\frac{4\pi}{3}\int_x^1
\tilde\varepsilon(u)\frac{d\tilde r^3(u)}{du}\,du.
\end{aligned}
\label{eq:spherical_reconstruction}
\end{equation}
This representation enforces $\tilde r(1)=\tilde m(1)=0$ and monotonicity of radius and enclosed mass. Numerically, both integrals use cumulative trapezoidal quadrature, with the mass integral evaluated in $\tilde r^3$ to integrate constant-density shells exactly.

Our DeepONet parametrizes the positive field as
\begin{equation}
\hat g_{\boldsymbol\theta}(x;\mathbf a)
=0.2\,\operatorname{softplus}\!
\left[\sum_{k=1}^{K}w_k b_k(\mathbf a)t_k(x)\right]
+10^{-8},
\end{equation}
where $b_k$ and $t_k$ are branch and trunk outputs, respectively, and $w_k$ are trainable weights of the bias-free output layer. The trunk receives $x$, while the branch receives
\begin{equation}
\mathbf a=
\left(
\{\tilde\varepsilon(x_i)\}_{i=0}^{N-1},
\{\tilde p(x_i)\}_{i=0}^{N-1},
h_c
\right),
\qquad x_i=\frac{i}{N-1}.
\end{equation}

Training minimizes the TOV residual evaluated on the reconstructed fields. In these coordinates, the required field and its regular central limit are
\begin{equation}
\begin{aligned}
\mathcal G(x)
&=\frac{2h_c\tilde r^2(\tilde r-2\tilde m)}
{\tilde m+4\pi\tilde r^3\tilde p},\\
g_c&=\frac{3h_c}{2\pi[1+3\tilde p(1)]}.
\end{aligned}
\end{equation}
The loss is
\begin{equation}
\begin{aligned}
\mathcal L_{\rm sph}
={}&\left\langle
\frac{1}{N-1}\sum_{i=0}^{N-2}
\left[
\frac{\hat g_{\boldsymbol\theta}(x_i)-\mathcal G(x_i)}
{\operatorname{sg}(|\mathcal G(x_i)|+0.1g_c)}
\right]^2
\right.\\
&\left.\hspace{12pt}
+\left[
\frac{\hat g_{\boldsymbol\theta}(1)-g_c}{g_c}
\right]^2
\right\rangle,
\end{aligned}
\end{equation}
where $\langle\cdot\rangle$ denotes an average over training stars and $\operatorname{sg}$ stops gradient propagation through the normalization. The first term includes the surface and excludes the centre, where the analytic regularity condition supplies the second term. No additional boundary penalties or labeled stellar solutions are required. The observables follow directly as $R=\tilde r(0)/\sqrt{\varepsilon_c}$ and $M=\tilde m(0)/\sqrt{\varepsilon_c}$. The training workflow is summarized in Algorithm~\ref{alg:pino-training-tov} and the loss-evolution is visualized in FIG.~\ref{fig:training}.

\begin{algorithm*}[t]
\caption{Physics-Informed DeepONet Training for Spherical Solutions}
\label{alg:pino-training-tov}
\begin{algorithmic}[1]
\Require $\mathcal D=\{(\mathcal E_i,\epsilon_{c,i},q_i)\}_{i=1}^{40000}$:
EoS curves, central energy densities, and axis ratios
\Require $\mathcal D_{\rm train},\mathcal D_{\rm val}$:
training and validation subsets of $\mathcal D$
\Require $\mathcal A$: candidate DeepONet architectures
\Require $(\textsc{Adam},\eta_0,\beta_1,\beta_2)$:
optimizer and associated parameters
\Require $\textsc{CosineScheduler}$:
cosine restarts with amplitude decay
\Require $\mathcal X=\{(s_j,\mu_j)\}_{j=1}^{N_{\rm grid}}$:
collocation grid
\Require $\mathcal L_{\rm physics}$:
main-text residual loss
\Require $N_{\rm epoch},P$:
maximum epochs and early-stopping patience
\Ensure Trained $\texttt{DeepONet}^{\star}$

\State $L^\star_{\rm val}\gets+\infty$
\For{$A\in\mathcal A$}
    \State Initialize $\texttt{DeepONet}_A$ with parameters $\Theta$
    \State Initialize Adam and cosine scheduler
    \State $L_{A,\rm best}\gets+\infty$; $n_{\rm plateau}\gets0$
    \For{$e=1$ to $N_{\rm epoch}$}
        \For{minibatch $\mathcal B\subset\mathcal D_{\rm train}$}
            \State $L\gets\Call{PhysicsLoss}{\Theta,\mathcal B,\mathcal X}$
            \State $\Theta\gets
            \textsc{AdamStep}(\Theta,\nabla_\Theta L,\eta,\beta_1,\beta_2)$
            \State Update $\eta$ using $\textsc{CosineScheduler}$
        \EndFor
        \State $L_{\rm val}\gets
        \Call{PhysicsLoss}{\Theta,\mathcal D_{\rm val},\mathcal X}$
        \If{$L_{\rm val}<L_{A,\rm best}$}
            \State $L_{A,\rm best}\gets L_{\rm val}$;
            $n_{\rm plateau}\gets0$
        \Else
            \State $n_{\rm plateau}\gets n_{\rm plateau}+1$
        \EndIf
        \If{$L_{\rm val}<L^\star_{\rm val}$}
            \State $L^\star_{\rm val}\gets L_{\rm val}$
            \State $\texttt{DeepONet}^{\star}\gets
            \textsc{Copy}(\texttt{DeepONet}_A,\Theta)$
        \EndIf
        \If{$n_{\rm plateau}\geq P$}
            \State \textbf{break}
        \EndIf
    \EndFor
\EndFor
\State \Return $\texttt{DeepONet}^{\star}$

\Function{PhysicsLoss}{$\Theta,\mathcal B,\mathcal X$}
    \For{$(\mathcal E_i,\epsilon_{c,i},q_i)\in\mathcal B$}
        \State Construct $h_{c,i}$ and branch input $\vec z_i$
        \State Evaluate $\texttt{DeepONet}_{A,\Theta}$ on $\mathcal X$
        \State Determine $r_e,\Omega,H,\epsilon,p,\alpha$
        from the predicted fields and $\mathcal E_i$
        \State Evaluate $S_u,A_u,K_{u,n}$ for
        $u\in\{\rho,\gamma,\omega\}$
        \State Compute $\bar u$ using Eq.~\eqref{eq:supp-green}
        \State $\ell_i\gets\mathcal L_{\rm physics}$
        evaluated from the predictions and integral-equation outputs
    \EndFor
    \State \Return $|\mathcal B|^{-1}\sum_{i\in\mathcal B}\ell_i$
\EndFunction
\end{algorithmic}
\end{algorithm*}
\begin{table}[t]
\centering
\caption{Real observational dataset. Masses and equatorial radii
are posterior medians with equal-tailed $90\%$ credible intervals,
using sample weights where supplied. Spin frequencies are treated
as fixed inputs. PSR J0952$-$0607 contributes only to the
maximum-mass constraint; its quoted mass uncertainty is retained
as supplied and is not converted to a $90\%$ interval.}
\label{tab:real-data}
\setlength{\tabcolsep}{3pt}
\resizebox{\columnwidth}{!}{%
\begin{tabular}{lccc}
\hline\hline
Source & $M\,[M_\odot]$ & $R_{\rm eq}\,[\mathrm{km}]$ & $\nu\,[\mathrm{Hz}]$ \\
\hline
PSR J0030$+$0451
& $1.289^{+0.288}_{-0.233}$ & $12.57^{+2.24}_{-2.00}$ & 205.53 \\
PSR J0740$+$6620
& $2.073^{+0.113}_{-0.113}$ & $12.49^{+2.26}_{-1.34}$ & 346.53 \\
PSR J0437$-$4715
& $1.418^{+0.057}_{-0.058}$ & $11.36^{+1.62}_{-0.99}$ & 173.69 \\
PSR J0614$-$3329
& $1.436^{+0.118}_{-0.116}$ & $11.16^{+2.86}_{-1.95}$ & 317.59 \\
PSR J1231$-$1411
& $1.060^{+0.112}_{-0.054}$ & $13.52^{+0.43}_{-0.90}$ & 271.45 \\
4U 1608$-$52
& $1.885^{+0.502}_{-0.470}$ & $9.81^{+4.43}_{-4.37}$ & 620.00 \\
KS 1731$-$260
& $1.601^{+0.596}_{-0.654}$ & $9.16^{+3.75}_{-3.85}$ & 524.00 \\
SAX J1748.9$-$2021
& $1.664^{+0.456}_{-0.484}$ & $10.20^{+3.80}_{-4.90}$ & 442.40 \\
4U 1636$-$536
& $1.775^{+0.322}_{-0.368}$ & $15.07^{+0.84}_{-2.04}$ & 581.91 \\
SRGA J144459.2$-$604207
& $1.917^{+0.532}_{-0.636}$ & $11.78^{+3.62}_{-2.50}$ & 447.87 \\
\hline
PSR J0952$-$0607
& $2.35\pm0.17$ & --- & 707.3 \\
\hline\hline
\end{tabular}%
}
\end{table}
\begin{figure*}
    \centering
    \includegraphics[width=0.48\textwidth]{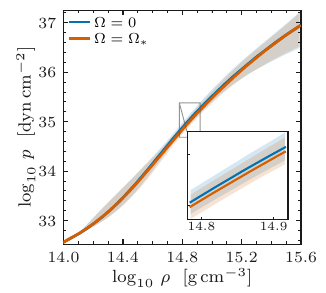}
    \includegraphics[height=0.43\textwidth]{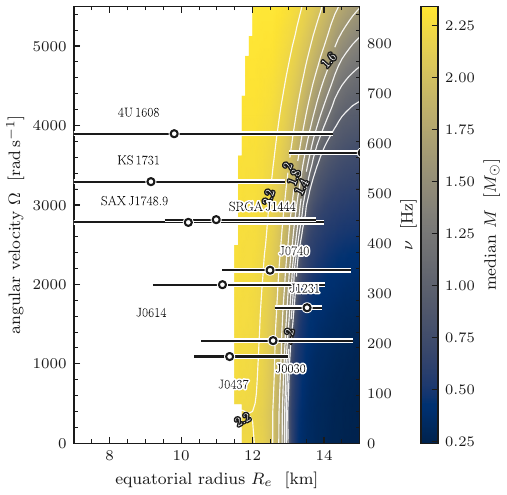}
    \includegraphics[width=0.48\textwidth]{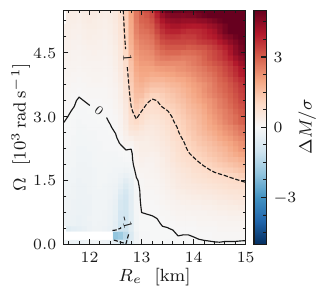}
    \includegraphics[width=0.48\textwidth]{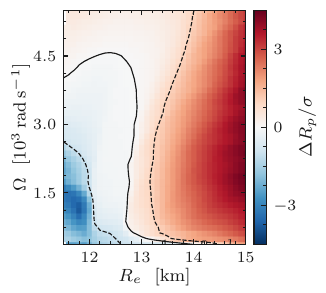}
    \caption{\label{fig:BHI2} Analysis of real data with rotaing and static solvers yield posterior credible intervals in the pressure-density plane~(\textit{top left}) and posterior predictive values of mass as a function of radius and and angular speed~(\textit{top}). Deviations between posterior predictive values of mass~(\textit{bottom left}) and polar radius~(\textit{bottom right})as a function of equatorial radius and angular speed between the rotation-informed and the static runs are also shown. It can be seen that incorporating rotation increases support for softer EoS models and larger values of polar radii for high equatorial radius~(much more modestly than the simulated dataset given the much larger measurement uncertainties).}
\end{figure*}
\begin{figure*}
    \centering
    
    \includegraphics[width=0.48\textwidth]{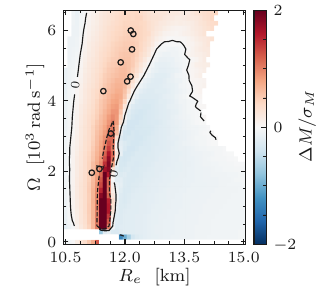}
    \includegraphics[width=0.48\textwidth]{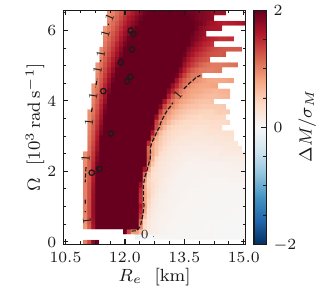}
    \includegraphics[width=0.48\textwidth]{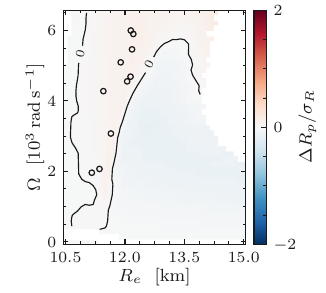}
    \includegraphics[width=0.48\textwidth]{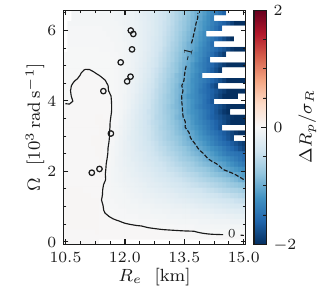}
    \caption{\label{fig:BHIMC2} 
    The PPD mass~(top) and polar-radius~(bottom) deviations from ground truth versus $(\Omega,R_e)$ for our~(center) and approximate~(right) solvers, all using simulated mass-radius measurements.
    }
\end{figure*}
\noindent

\textbf{Data and Additional Analysis Results} --- We combine ten independent mass-radius measurements spanning rotation-powered and accreting MSPs. NICER pulse-profile measurements comprise PSRs J0030+0451, J0740+6620, J0437$-$4715, J0614$-$3329, and J1231$-$1411~\cite{Riley2019,Salmi2024J0740,Choudhury2024,Miller2026J0614,Salmi2024J1231}. Accreting sources include photospheric-radius-expansion burst measurements of 4U 1608$-$52, KS 1731$-$260, and SAX J1748.9$-$2021~\cite{Guver2010,Ozel2012,GuverOzel2013,Steiner2013,Ozel2016}, superburst pulse-profile measurements of 4U 1636$-$536~\cite{Kini2025}, and burst-ensemble modeling of SRGA J144459.2$-$604207~\cite{Galloway2026}. We account for the source-specific radius priors while retaining independent radio-timing mass constraints, and additionally incorporate the mass measurement of PSR J0952$-$0607 as a constraint on the maximum supported mass at its observed spin~\cite{Romani2022,Bassa2017}. 
For simulated observations, we generate thirty APR4\_EPP~\citep{Akmal1998, Read2009} stars with spin frequencies drawn uniformly between $300$ and $1000\,\mathrm{Hz}$ and masses drawn from the Galactic double-Gaussian distribution~\cite{Alsing2018}, truncated to the available stable mass range at each spin; their equatorial radii are computed independently of our neural solvers by interpolating \texttt{RNS} equilibrium sequences~\cite{StergioulasFriedman1995}. The mass-radius-spin measurements for each observation are summarized in Table~\ref{tab:real-data}. Each simulated measurement comprises $4000$ samples from an uncorrelated bivariate Gaussian centered on the injected mass and radius, with $\sigma_M=0.05M$ and $\sigma_R=0.05R_{\rm e}$, truncated at $M>0.2\,M_\odot$ and $R_{\rm e}>3\,\mathrm{km}$. The sample size and fractional uncertainties represent an optimistic scenario motivated by eXTP forecasts~\cite{Li2025eXTP}.

We now present additional results from the analyses summarized in the main text. In FIG.~\ref{fig:BHI2}, we show the pressure-density posterior and posterior predictive masses and polar radii for the analysis with real events. In FIG.~\ref{fig:BHIMC2}, we show the fractional deviation of the posterior predictive masses from the static and uniformly rotating analyses relative to the ground truth for the simulated events. 
\end{document}